\documentclass[pra,twocolumn,superscriptaddress,amsmath,amssymb,showpacs,showkeys]{revtex4}

\usepackage{graphicx}
\usepackage[dvipsnames]{color}
\usepackage[colorlinks=true,citecolor=myblue,linkcolor=myblue,urlcolor=myblue]{hyperref}
\definecolor{myblue}{named}{MidnightBlue}
\definecolor{mygray}{gray}{0.35}
\usepackage[SquareTraceBrackets]{quantum}
\usepackage{bm,natbib,upgreek,amsbsy,mathrsfs,amsfonts}

\usepackage[export]{adjustbox}%
\usepackage{pifont}

\DeclareMathOperator{\cov}{\textbf{C}}
\DeclareMathOperator{\covf}{\textbf{\textsf{\small Cov}}}
\DeclareMathOperator{\varf}{\textbf{\textsf{\small Var}}}
\newcommand{\mpi}{-1}
\newcommand{\transpose}{\!\reflectbox{\rotatebox[origin=c]{180}{\small $\perp$}}}

\newcommand{\cone}{\textcolor{mygray}{\ding{202}}}
\newcommand{\ctwo}{\textcolor{mygray}{\ding{203}}}
\newcommand{\cthree}{\textcolor{mygray}{\ding{204}}}
\newcommand{\cfour}{\textcolor{mygray}{\ding{205}}}

\begin{document}

\title{\textsf{The role of entanglement in calibrating optical quantum gyroscopes}} 

\author{Pieter Kok}\email{p.kok@sheffield.ac.uk}
\affiliation{Department of Physics and Astronomy, University of Sheffield, Sheffield S3 7RH, United Kingdom}

\author{Jacob Dunningham}
\affiliation{Department of Physics and Astronomy, University of Sussex, Brighton BN1 9QH, United Kingdom}

\author{Jason F. Ralph}
\affiliation{Department of Electrical Engineering and Electronics, The University of Liverpool, Brownlow Hill, Liverpool, L69 3GJ, United Kingdom.}

\begin{abstract} 
\noindent
We consider the calibration of an optical quantum gyroscope by modeling two Sagnac interferometers, mounted approximately at right angles to each other. Reliable operation requires that we know the angle between the interferometers with high precision, and we show that a procedure akin to multi-position testing in inertial navigation systems can be generalized to the case of quantum interferometry. We find that while entanglement is a key resource within an individual Sagnac interferometer, its presence between the interferometers is a far more complicated story. The optimum level of entanglement depends strongly on the sought parameter values, and small but significant improvements may be gained from choosing states with the optimal amount of entanglement between the interferometers.  
\end{abstract}

\date{\today}

\maketitle

\section{Introduction}\noindent
Quantum metrology and quantum parameter estimation offer great potential improvements in precision measurement. Recent experiments have demonstrated quantum improvements in measuring protein concentration \cite{crespi12}, tracking lipid granules in yeast cells \cite{taylor13}, and searching for gravitational waves \cite{LIGO11}. In optical systems, the standard way to frame problems in quantum metrology is as a measurement of the phase of an optical signal. The aim is to improve the precision of such measurements from the classical shot noise limit (SNL) to the quantum mechanical Heisenberg limit (HL) \cite{giovannetti11}. It has been recognized that any practical implementation of quantum metrology requires methods to deal with effects due to environmental noise and dissipation \cite{tsang13}. Quantum error correction has been proposed to combat the effect of noise \cite{Dur14,Arrad14,Kessler14}, and loss-tolerant metrology protocols have been designed and implemented to address some of the negative effects of dissipation \cite{pezze08,Kacprowicz10,xiang11,Marino12}. It has been shown that the measurement of $d$ phases in an interferometer can obtain an improvement of a factor $O(d)$ in the precision when multi-mode entanglement is used \cite{datta13}. This behavior persists in the presence of photon loss \cite{yue14}, even though multi-mode entanglement is highly susceptible to such processes \cite{knott14}. When the loss parameters are also estimated, there is a trade-off between the attainable precision of the phase estimation and the estimation of these parameters \cite{crowley14}. However, loss and noise are not the only causes for imperfect metrology. The accuracy of a composite sensor system is only partially determined by the precision of the individual measurements. Other sources of imperfection can include badly characterized responses to non-standard stimuli, or couplings between the parameters of interest. The performance of any larger scale system---i.e., one containing a number of individual sensors---will be limited by the presence of such nuisance parameters, but this aspect of quantum metrology has been somewhat overlooked. 

In this paper, we address the problem of nuisance parameters arising from unwanted couplings between sensors in practical quantum parameter estimation. Such couplings affect the measurement precision---defined by the mean square error (MSE)---and must also be estimated, even if we are ultimately not interested in their numerical value \cite{kay93}. For a single parameter, the quantum Cram\'er-Rao bound (QCRB) puts a lower limit on the MSE, determined by the inverse of the quantum Fisher information (QFI) \cite{helstrom69,braunstein94}. Multiple parameters lead to a QFI matrix, the inverse of which provides lower bounds for the MSE covariance matrix \cite{yuen73,helstrom74}. Nuisance parameters are part of this multi-parameter estimation problem. While the QCRB for a single parameter can generally be attained, this is not always true of the QCRB for multiple parameters \cite{paris09,gill13}. Where multiple parameters are being estimated, it matters whether the generators of translation of the parameters commute or not, with implications for the optimal strategies of the parameter estimation procedures \cite{Macchiavello03,ballester04,ballester04b,imai07,genoni13,gao14}. Even though multi-mode entanglement can be used to improve the estimation of multiple phase parameters beyond the classical SNL \cite{datta13}, this is not always the case. For the example considered in this paper, we show that the optimum entanglement is a function of the nuisance parameters being estimated and that---for a range of parameter values with practical relevance---the presence of entanglement can be detrimental to the estimation process.
\begin{figure}[t]
\includegraphics[width=7cm]{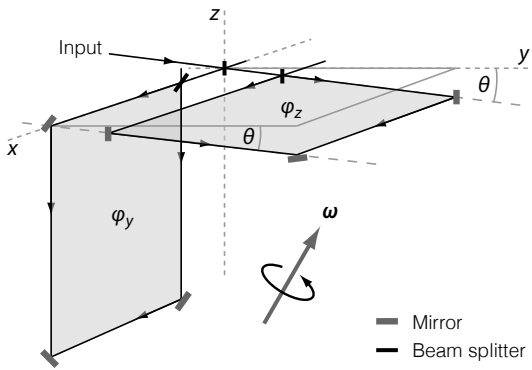}
\caption{Two nearly orthogonal (coupled) Sagnac interferometers that are misaligned by an angle $\theta$. The entire system rotates with angular velocity $\bm\omega$, and the resulting phase
shifts $\varphi_y$ and $\varphi_z$ in the interferometers can be used to estimate $\bm\omega$. For clarity, the third interferometers measuring $\varphi_x$ and the photodetectors are omitted.}
\label{fig:sagnac}
\end{figure}

We consider a simplified optical gyroscope configuration based on two Sagnac interferometers, shown in Fig.~\ref{fig:sagnac}, in which the two (nominally) orthogonal interferometers are misaligned by a small angle $\theta$. We find that in such circumstances entanglement can hinder the calibration of the misalignment. In fact, we will show that entanglement (or classical correlations) in the quantum state shared by the two interferometers can limit the precision of the estimation process for $\theta$, whilst entanglement can assist in the determination of the phases of interest, $\varphi_y$ and $\varphi_z$. These results are valid for a wide range of physically relevant parameter values, and we indicate how the calibration process can be generalized to a set of three Sagnac interferometers measuring arbitrary three dimensional rotation rates. 

The misalignment of gyroscopic sensors is a well-known problem in the construction of inertial navigation systems~\cite{savage98,aggarwal08}. Fibre-optical Sagnac interferometers are often used in modern ``strapdown'' inertial navigation systems (i.e. fixed sensors within the body of the navigation system)~\cite{lefevre93}. In such systems, three fibre-optical gyroscopes (FOGs) and three accelerometers are mounted in the inertial measurement unit of the navigation system. The gyroscopes provide measurements of the rotation rates about their axes, where the axes are normally designed to form an orthogonal triad. Integrating the rotation rates provides estimates of the angles of rotation of the system, relative to a set of reference axes. The angles are used to determine the system's orientation, but they are also used to resolve the measured accelerations into the reference axes to determine the system's translational motion (i.e. velocity and position). As a result, the accuracy of the gyroscopic sensors is often a limiting factor in the overall performance of an inertial navigation system. The gyroscopes will have mis-alignment errors due to mechanical tolerances in their construction and systematic errors in the measurement devices, both of which limit the accuracy of the sensors. Calibrating these errors, and correcting for them in software, is one way to improve the accuracy of the inertial navigation system, and this has become standard practice in many applications~\cite{savage98,aggarwal08}. After production, an inertial measurement unit will undergo a ``multi-position'' test. It is rotated through a set of known rotations, using a very accurate reference system, to obtain a static measurement value and then subjecting the unit to a known rotation rate after each rotation---normally, at least six different rotations/orientations are used to calibrate non-orthogonality within the triads of sensors, static bias measurement errors in each sensor, and scaling errors in the measurement of the known rotation rates~\cite{savage98,aggarwal08}. In this paper, we are primarily interested in the example where there is coupling between two non-orthogonal gyroscopes, measuring rotation rates about the $y$- and $z$-axes, so we will consider the simplest of these calibration processes, the measurement of a fixed (but otherwise unknown) rotation rate, followed by another measurement after a rotation by $\pi/2$ about the $x$-axis---although we will also indicate how this may be extended to deal with a triad of three gyroscopes.

\section{Coupled Sagnac Interferometers}\noindent
The Sagnac interferometer \cite{tartaglia15} can be described quantum mechanically in a very similar way to the Mach-Zehnder interferometer, but instead of two spatially different paths in the latter, the Sagnac interferometer has a single loop with two counter-propagating modes, $a$ and $b$. The phase shift induced by a rotation of the interferometer can be written as a unitary transformation
\begin{align}\label{eq:sagnacU}
 U (t) = \exp\left[ -i \bm{\omega}\cdot\mathbf{e} \left( \hat{n}_a - \hat{n}_b \right) t \right]\, ,
\end{align}
where $\bm{\omega}$ is a normalized rotation rate, $\mathbf{e}$ is the normal vector to the plane of the interferometer, and $\hat{n}_a$, $\hat{n}_b$ are the number operators in modes $a$ and $b$. (The Sagnac phase shift is dependent on a number of device-specific parameters---including operating wavelength, path length and enclosed area~\cite{lefevre93} ---and it is proportional to the angular velocity vector applied to the interferometer so we will use a normalized rotation rate to remove to the explicit dependence on these parameters and to simplify the presentation of the results below). For simplicity, we assume that the Sagnac interferometer lies entirely in the $xy$-plane ($\mathbf{e} = \hat{\rm{e}}_z$), and we define $\varphi_j \equiv \omega_j t$ and $\hat{n} \equiv \hat{n}_a - \hat{n}_b$. Then we can write the transformation in Eq.~(\ref{eq:sagnacU}) as $U (\varphi_z) = \exp\left[ -i \varphi_z \hat{n} \right]$. In the usual notation where $U = \exp(-iHt/\hbar)$, the Hamiltonian becomes $H = \hbar \omega_z \hat{n}$ with $t$ the interaction time (assumed to be known with arbitrary precision), and we will now set $\hbar=1$. Clearly, measurements of the phase $\varphi_z$ can be used to determine the rotation rate $\omega_z$ applied to the gyroscope. We can construct a second Sagnac interferometer in the $xz$-plane ($\mathbf{e} = \hat{\rm{e}}_y$) to determine the rotation rate $\omega_y$, and a third can be added to determine $\omega_x$. The use of three such gyroscopes allows a general rotation rate about an arbitrary axis to be determined \cite{culshaw06}. We concentrate on the case with two interferometers for clarity, but the generalisation to three interferometers will also be discussed below. 

In any practical construction, the two Sagnac interferometers will not be perfectly perpendicular (and when the interferometer is constructed from optical fibres it may not lie perfectly in a plane). Let $\hat{n}_y$ be the number difference operator for the counter-propagating modes of the interferometer in the $xz$-plane, and $\hat{n}_z$ the equivalent operator for the interferometer in the $xy$-plane. Furthermore, let $\theta$ be the angle with which the $\varphi_z$ interferometer is misaligned, shown in Fig.~\ref{fig:sagnac}: 
\begin{align}\label{eq:zprime}
 \hat{\rm{e}}_z' = \cos\theta\, \hat{\rm{e}}_z + \sin\theta\, \hat{\rm{e}}_y\, . 
\end{align}
The transformation of the optical state inside the interferometers then becomes
\begin{align}\label{eq:sagnacUcoupled}
 U (\bm{\varphi}) = \exp\left[ -i \left(\varphi_y \hat{n}_y + \cos\theta\, \varphi_z \hat{n}_z + \sin\theta\, \varphi_y \hat{n}_z \right)  \right]\, ,
\end{align}
leading to a Hamiltonian for the system 
\begin{align}
 H = \omega_y \hat{n}_y + \cos\theta\; \omega_z \hat{n}_z + \sin\theta\; \omega_y \hat{n}_z\, . 
\end{align}
There is now a coupling between the two interferometers given by the term $ \sin\theta\, \omega_y \hat{n}_z$. As a consequence, we have three unknown parameters, $\varphi_y$, $\varphi_z$, and $\theta$, but we measure only two observables, $\hat{n}_y$ and $\hat{n}_z$. The problem is therefore underdetermined, and we cannot extract the true values of $\varphi_y$, $\varphi_z$ without an unknown bias.

\begin{figure}[t]
\includegraphics[width=4.3cm]{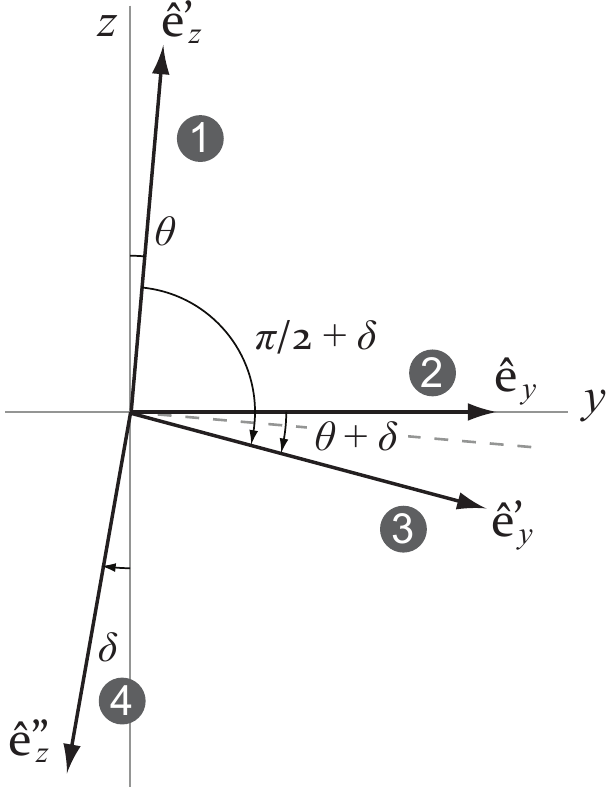}
\caption{The coordinate system and the normal vectors to the Sagnac interferometers. Gyroscopes \cone\ and \ctwo\ take one set of measurements, and after a rotation of $\pi/2+\delta$, gyroscopes \cthree\ and \cfour\ take another set of measurements. Gyroscope \cone\ rotates to \cthree, and \ctwo\ rotates to \cfour.}
\label{fig:coords}
\end{figure}

To remedy this, we may rotate the system of gyroscopes by $\pi/2$, whilst keeping the applied rotation rate fixed with respect to an external reference system. This allows us to measure four different observables (two for each orientation), given our three unknown parameters. Assuming that the system is rigid, $\theta$ remains unchanged. However, we do need to introduce a new nuisance parameter $\delta$ that encodes imperfections in the $\pi/2$ rotation. This leaves us with four parameters and four observables. We will show in Sec.~\ref{sec:qfi} that this leads to a linearly independent set of four estimators, but first we establish the coordinate system local to our gyroscopes, shown in Fig.~\ref{fig:coords}. In the original position of Fig.~\ref{fig:sagnac}, the $y$-rotation \ctwo\ is matched to the local $y$-axis by definition, while the $z$-rotation \cone\ is misaligned according to Eq.~\eqref{eq:zprime}. After a rotation of $\pi/2 + \delta$ about the sensor $x$-axis, gyroscope \cthree\ is now aligned along the normal $\hat{\rm{e}}_y'$ with 
\begin{align}\label{eq:yprime}
 \hat{\rm{e}}_y' = \cos(\theta+\delta)\, \hat{\rm{e}}_y - \sin(\theta+\delta)\, \hat{\rm{e}}_z\, ,
\end{align}
and gyroscope \cfour\ is aligned along the normal $\hat{\rm{e}}_z''$ with 
\begin{align}\label{eq:yprime}
 \hat{\rm{e}}_z'' = -\cos\delta\, \hat{\rm{e}}_z - \sin\delta\, \hat{\rm{e}}_y\, .
\end{align}
The measured phases are decomposed in the same way as the normal vectors ($\varphi_j \leftrightarrow \hat{\rm {e}}_j$). In the ideal case where $\theta = \delta = 0$ we have $\varphi_y = \varphi_y'$, and $\varphi_z = \varphi_z' = -\varphi_z''$.

Let the joint state in the two nearly perpendicular gyroscopes be denoted by $\rho_{ij}$, where $i$ and $j$ indicate the gyroscopes \cone--\cfour\ in Fig.~\ref{fig:coords}. The two gyroscopic measurements are equivalent to a single measurement with four gyroscopes simultaneously, with a joint state $\rho \equiv \rho_{12} \otimes \rho_{34}$. In general, the optimal state $\rho_{12}$ for \cone\ and \ctwo\ will not be the same as the optimal state $\rho_{34}$ for \cthree\ and \cfour, since the evolution $U_{12}$ of the gyroscopes \cone\ and \ctwo\ is not equal to the evolution $U_{34}$ for the gyroscopes \cthree\ and \cfour\ due to the different relative rotation $\bm\omega$. The transformation of the optical state due to the rotation rate $\bm{\omega}$ can be written on the joint system as $U \equiv U_{12} U_{34}$, with
\begin{align}
 U_{12} & = \exp\left[ -i\left( \varphi_z' \hat{n}_z + \varphi_y \hat{n}_y \right) \right]\\
 U_{34} & = \exp\left[ -i\left( \varphi_y' \hat{n}_z' + \varphi_z'' \hat{n}_y' \right) \right]\, ,
\end{align}
which depend in a nontrivial way on the parameters $\bm{\varphi}$.
The four operators $\hat{n}_y$, $\hat{n}_y'$, $\hat{n}_z$, and $\hat{n}_z'$ commute, and can be measured simultaneously. The joint evolution then becomes
\begin{align}\label{eq:unitary}
 U(\bm{\varphi}) & =  \exp\left\{ -i\left[ \phi_y \hat{n}_y - \beta\left(\theta+\frac{\pi}{2}\right) \hat{n}_z \right] \right\} \cr 
 & \qquad \times  \exp\left\{ -i\left[ \beta\left(\delta+\frac{\pi}{2}\right) \hat{n}_y' + \beta(\theta+\delta) \hat{n}_z'  \right] \right\} ,
\end{align}
where
\begin{align}
 \beta(\alpha) = \phi_y \cos\alpha - \phi_z \sin\alpha\, .
\end{align}
and $\bm{\varphi} \equiv (\varphi_y,\varphi_z,\theta,\delta)$. The evolution $U$ in Eq.~\eqref{eq:unitary} is expressed entirely in terms of the measurable observables $\hat{n}_y$, $\hat{n}_y'$, $\hat{n}_z$, and $\hat{n}_z'$, and the four unknown parameters $\bm{\varphi}$. In the next section we use this evolution to calculate the quantum Fisher information and the Cram\'er-Rao bound for these parameters.

\section{Covariance and Fisher Information}\label{sec:qfi}\noindent
To determine the ultimate precision with which we can estimate the Sagnac phases and the couplings between them, we consider the quantum Cram\'er-Rao bound
\begin{align}\label{eq:crb}
 \covf(\bm{\varphi}) \geq \frac{1}{N}\; \mathbf{I}_{Q}^{-1}(\bm{\varphi})\, , 
\end{align}
where $\covf(\bm{\varphi})$ is the covariance matrix of the four variables $\bm{\varphi}$, $N$ is the number of independent measurements, and $\mathbf{I}_{Q}(\bm{\varphi})$ is the quantum Fisher information (QFI) matrix of the three variables \cite{helstrom69,gammelmark14} with elements:
\begin{align}\label{eq:qfim}
 [ \mathbf{I}_{Q}(\bm{\varphi})]_{ij} = 2 \partial_i \partial_{\tilde{j}} \log {\Abs{\Braket{\psi(\bm{\varphi})|\psi({\tilde{\bm{\varphi}}})}}}^2_{\tilde{\bm{\varphi}}=\bm{\varphi}} \, ,
\end{align}
where $\partial_i$ is the derivative with respect to $\varphi_i$, and $\partial_{\tilde{j}}$ the derivative with respect to $\tilde{\varphi}_j$. Let $\mathbf{G}=(G_y, G_z, G_\theta, G_\delta)^{\transpose}$ be the tuple of generators of translation in our four parameters. Generally, a generator of translation $G_\alpha$ of a parameter $\alpha$ can be defined as \cite{pang14} 
\begin{align}\label{eq:gen}
 G_\alpha \equiv i U^\dagger \partial_\alpha U\, . 
\end{align}
This allows us to relate the derivative of the quantum state $\ket{\psi}$ with respect to $\varphi_i$ to the generator $G_i$ via a Taylor expansion of $U$. Evaluating the matrix elements of the QFI matrix for pure states $\ket{\psi}$ then yields 
\begin{align}\label{eq:qfi}
  [\mathbf{I}_{Q}(\bm{\varphi})]_{ij} & =4 \left( \frac12\Braket{\{ G_i,G_j\}}_{\bm{\varphi}} - \braket{G_i}_{\bm{\varphi}}\braket{G_j}_{\bm{\varphi}} \right) \cr & \equiv 4 [\cov_S(\bm{\mathbf{G}})]_{ij}\, ,
\end{align}
where  $\braket{O}_{\bm{\varphi}} \equiv \braket{\psi(\bm{\varphi})|O|\psi(\bm{\varphi})}$ for some operator $O$, and $[\cov_S(\mathbf{G})]_{ij}$ is the symmetrized covariance matrix element between operators $G_i$ and $G_j$, originating from the fact that the quantum Fisher information matrix in equation (\ref{eq:qfim}) is derived from the symmetric logarithmic derivative \cite{helstrom69}. Since all our generators commute with each other, we can ignore this technical requirement and drop the subscript $S$.

Applying Eq.~\eqref{eq:gen} to $U (\bm{\varphi})$ in Eq.~(\ref{eq:unitary}) for the parameters $\varphi_y$, $\varphi_z$, $\theta$ and $\delta$, we obtain the generators
\begin{align}\label{eq:generators}
 G_y & = \hat{n}_y + \sin\theta\; \hat{n}_z - \sin\delta\; \hat{n}_y' + \cos(\theta+\delta)\; \hat{n}_z'\, , \cr
 G_z & = \cos\theta\; \hat{n}_z - \cos\delta\; \hat{n}_y' - \sin(\theta+\delta)\; \hat{n}_z'\, , \cr
 G_{\theta} & =  \beta(\theta)\; \hat{n}_z + \beta\left(\theta+\delta+\frac{\pi}{2}\right)\; \hat{n}_z'\, , \cr
 G_{\delta} & = - \beta(\delta)\; \hat{n}_y' + \beta\left(\theta+\delta+\frac{\pi}{2}\right)\; \hat{n}_z'\, .
\end{align}
The relation between the generators $G_j$ and the observables $\hat{n}_k$ is linear and can be expressed in matrix form as $\mathbf{G} = M\hat{\mathbf{n}}$ with 
\begin{align}
 M =
 \begin{pmatrix} 
  1 & \sin\theta & \cos(\theta+\delta) & -\sin\delta \cr
  0 & \cos\theta & \sin(\theta+\delta) & -\cos\delta \cr
  0 & \beta(\theta) & \beta(\theta+\delta+\pi/2) & 0 \cr
  0 & 0 &  \beta(\theta+\delta+\pi/2) & -\beta(\delta)
 \end{pmatrix} ,
\end{align}
and $\hat{\mathbf{n}} = (\hat{n}_y,\hat{n}_z,\hat{n}_y',\hat{n}_z')^{\transpose}$. The determinant of this matrix is
\begin{align}
 \det M & = \left[ \beta(\theta) + \beta(\delta) \right] \beta\left(\theta+\delta+\frac{\pi}{2}\right) \cos(\theta) \cr 
 &\qquad  - \beta(\theta) \beta(\delta) \sin(\theta+\delta) \, ,
\end{align}
which is nonzero for most values of $\bm{\varphi}$, and in particular for our case of interest of small values of $\theta$ and $\delta$. The four observables can therefore be used to determine the four parameters unambiguously.

We calculate the QFI in terms of the matrix $M$:
\begin{align}
 [\mathbf{I}_{Q}(\bm{\varphi})]_{ij} & = 4 \left[ \braket{G_i G_j} - \braket{G_i} \braket{G_j} \right] \cr
  & = 4 \sum_{kl} M_{ik} M_{jl} [\cov(\hat{\mathbf{n}})]_{kl} \, ,
\end{align}
or 
\begin{align}\label{eq:qficompact}
 \mathbf{I}_{Q}(\bm{\varphi}) = 4 M\! \cov(\hat{\mathbf{n}}) M^{\,\transpose} \, ,
\end{align}
where $[\cov(\hat{\mathbf{n}})]_{kl} = \braket{\hat{n}_k\hat{n}_l} - \braket{\hat{n}_k}\braket{\hat{n}_l}$ are covariances that depend only on the state inside the interferometers (all the alignment information is encoded in $M$). This expression contains a large number of variables in $\cov(\hat{\mathbf{n}})$, corresponding to the extensive freedom to choose input states of the interferometers. However, we can drastically reduce the number of variables using simple symmetry arguments: Since the state $\rho$ is a tensor product $\rho_{12} \otimes \rho_{34}$, the $2\times 2$ off-diagonal sub matrices of $\cov(\hat{\mathbf{n}})$ are zero due to the fact that for this case $\braket{\hat{n}_k\hat{n}_l} = \braket{\hat{n}_k}\braket{\hat{n}_l}$. Moreover, to keep the resources in the gyroscopes identical between rotations, we take the photon number differences $(\Delta \hat{n}_y)^2$ and $(\Delta \hat{n}_z)^2$ the same in the gyroscope setting \cone+\ctwo\ and  \cthree+\cfour. This leads to the covariance matrix
\begin{align}\label{eq:covnum}
 \cov(\hat{\mathbf{n}}) = 
 \begin{pmatrix} 
  (\Delta \hat{n}_y)^2 & C_{yz}^{(12)} & 0 & 0 \cr
  C_{yz}^{(12)} & (\Delta \hat{n}_z)^2 & 0 & 0 \cr
  0 & 0 & (\Delta \hat{n}_y)^2 & C_{yz}^{(34)}  \cr
  0 & 0 & C_{yz}^{(34)} &(\Delta \hat{n}_z)^2
 \end{pmatrix} ,
\end{align}
where $C_{yz}^{(ij)} = \braket{\hat{n}_y\hat{n}_z} - \braket{\hat{n}_y}\braket{\hat{n}_z}$ for gyroscope pair $i$ and $j$. A priori there is no reason to choose different probe states for the two orientations \cone+\ctwo\ and  \cthree+\cfour\ during normal operation. However, during the calibration stage of the gyroscopes the rotation $\bm\omega$ is a precisely known rotation in an external reference frame, and its value will generally determine different optimal states $\rho_{12}$ and $\rho_{34}$. Here we choose $\rho_{12}=\rho_{34}$ to keep the analysis tractable. The optimal strategy for different $\rho_{12}$ and $\rho_{34}$, as well as the optimal rotation direction $\bm\omega$ will be the subject of future work. 

Without prior knowledge of $\theta$ and $\delta$ there is no reason to require different values for $(\Delta \hat{n}_y)^2$ and $(\Delta \hat{n}_z)^2$. We can therefore take 
\begin{align}
 (\Delta \hat{n}_y)^2 = (\Delta \hat{n}_z)^2 \equiv (\Delta \hat{n})^2 
\end{align}
and 
\begin{align}
 C_{yz}^{(ij)} = \lambda (\Delta \hat{n})^2\, , 
\end{align}
with $-1\leq \lambda \leq 1$ the correlation coefficient. Note that the correlation in this context means entanglement, since the optimal states are pure states \cite{braunstein94}, and classical correlations require mixed states.

\begin{figure}[t]
\includegraphics[width=8.5cm]{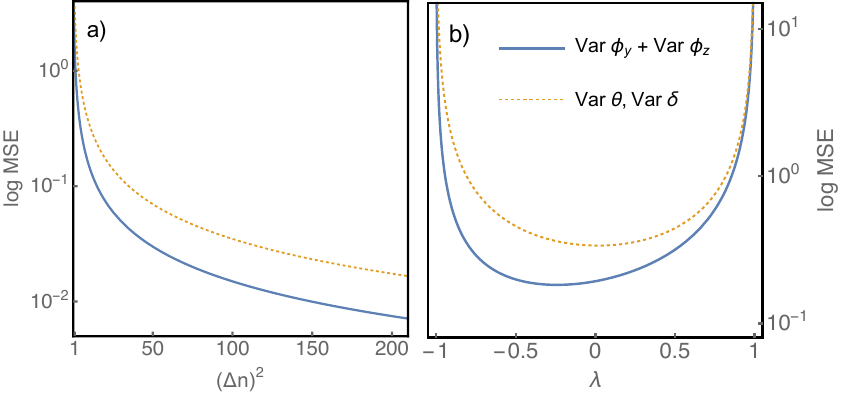}
\caption{Color online. The MSEs for $\varphi_y$, $\varphi_z$, $\theta$ and $\delta$ as a function of (a) the variance in the photon number $(\Delta \hat{n})^2$ in each Sagnac interferometer---assuming all four interferometers having the same $(\Delta \hat{n})^2$ with $\lambda=0$, and (b) the correlation coefficient $\lambda$ between photon number difference in two Sagnac interferometers. Here, $\theta = 0.02$ rad, $\delta = 0.013$ rad, $\varphi_y = 0.66$ rad,  $\varphi_z = 0.17$ rad,  and $(\Delta \hat{n})^2 =10$. The curve legend for the variances of $\varphi_y$, $\varphi_z$, $\theta$ and $\delta$ applies to both figures.}
\label{fig:crb}
\end{figure}

The diagonal elements of $\covf(\bm{\varphi})$ are the variances of the parameters of interest, namely $\varf\varphi_y$, $\varf\varphi_z$, and the nuisance parameters $\varf\theta$ and $\varf\delta$. We can choose to optimize any one of these variances, two or three of them, or all four. In the latter case, we need to choose a quantum state that minimizes 
\begin{align}\label{eq:covtrace}
 \Tr{\covf(\bm{\varphi})} \geq \frac{1}{N} \Tr{ \mathbf{I}^{-1}_{Q}(\bm{\varphi})}\, .
\end{align}
The right-hand side of Eq.~(\ref{eq:covtrace}) provides a bound on the optimal joint estimation of $\bm{\varphi}$ that may be achieved in the asymptotic limit of large $N$. Another interesting optimisation is to minimize the combination $\varf\varphi_y+\varf\varphi_z$. The values of the nuisance parameters $\theta$ and $\delta$ are only interesting in as far as they can be used to improve the accuracy of the overall sensor. They do not convey information about the rotation of the gyroscopes per se, but---once calibrated---they can be used to correct fixed errors due to couplings between the measurements of the interferometers. The MSE for a typical case of $\theta = 0.02$ rad, $\delta = 0.013$ rad, $\varphi_y = 0.66$ rad , and $\varphi_z = 0.17$ rad is shown in Fig.~\ref{fig:crb}.

\subsection{Variances for well-aligned gyroscopes}\noindent
The QFI in Eq.~\eqref{eq:qficompact} based on the generators in Eq.~(\ref{eq:generators}) and the covariance matrix in Eq.~(\ref{eq:covnum}) is easily calculated. The diagonal elements of its inverse can be written as
\begin{align}\label{eq:qfifinal}
 \left[\mathbf{I}^{\mpi}_{Q}(\bm{\varphi})\right]_{kk}  = \frac{1}{4[(\Delta \hat{n})^2 [\det M(\bm{\varphi})]^2} \frac{F_k(\bm{\varphi},\lambda)}{(1-\lambda^2) }\, , 
\end{align}
where the functions $F_k$ are lengthy expressions in $\bm{\varphi}$ and $\lambda$ (given in full in Appendix \ref{app:F}). Note that the diagonal elements of the inverse QFI are all proportional to $(\Delta \hat{n})^{-2}$, as well as the squared determinant of $M$. This confirms our intuition that we should maximize the variance $(\Delta \hat{n})^{2}$ in each Sagnac interferometer (for example using NOON states \cite{kok10}). 

We can also investigate the dependence of $[\mathbf{I}^{\mpi}_{Q}(\bm{\varphi})]_{kk}$ on the correlations in the states between the interferometers, $\lambda$. Consider first a perfect gyroscope with $\theta = \delta = 0$. Solving the simpler problem of only two generators $\hat{n}_y$ and $\hat{n}_z$ and ignoring the nuisance parameters entirely we obtain
\begin{align}\label{eq:ideal}
 \varf \varphi_{y} & \geq \frac{1}{4N} \frac{(\Delta\hat{n}_z)^2}{(\Delta\hat{n}_y)^2 (\Delta\hat{n}_z)^2-C_{\text{yz}}^2}\, , \cr
 \varf \varphi_{z} & \geq \frac{1}{4N} \frac{(\Delta\hat{n}_y)^2}{(\Delta\hat{n}_y)^2 (\Delta\hat{n}_z)^2-C_{\text{yz}}^2}\, .
\end{align}
On the other hand, taking the limit of $\theta , \delta \to 0$ in the four-parameter problem and setting $\kappa\equiv \varphi_y / \varphi_z$ yields a different result for $\varf\varphi_z$:
\begin{align}\label{eq:noisy}
 \varf \varphi_{z} & \geq \frac{1}{8N}  \frac{(\Delta\hat{n}_y)^2+\kappa^2(\Delta\hat{n}_z)^2+2\kappa C_{yz} }{(\Delta\hat{n}_y)^2 (\Delta\hat{n}_z)^2 -C_{yz}^2} ,
\end{align}
and this reduces to Eq.~\eqref{eq:ideal} only in the limit where $\kappa\to 0$ (up to an overall factor of two, since Eq.~\eqref{eq:noisy} contains two independent sets of measurements---initial and rotated by $\pi/2$---in this limit). The reason for the discrepancy between Eq.~(\ref{eq:ideal}) and Eq.~(\ref{eq:noisy}) is that the former does not take into account the nuisance parameters at all, while the latter takes into account the uncertainty in the nuisance parameters $\theta$ and $\delta$ when their mean value is equal to zero.

When we consider the mean square errors in Eq.~\eqref{eq:ideal}, we see that the right-hand side is minimized when $C_{yz}=0$. Since these are effectively two separate single-parameter measurements, the quantum Cram\'er-Rao bound is attainable asymptotically, and we conclude that, in the absence of alignment errors, any correlation between the input states in the two Sagnac interferometers---including entanglement---is detrimental to the precision of the optical quantum gyroscope.

\begin{table}[t]
\begin{ruledtabular}
\begin{tabular}{rrccr}
  $\varphi_y$~ & $\varphi_z$~ &  $\varf\varphi_y + \varf\varphi_z$ & $\varf\varphi_y + \varf\varphi_z$ & $\lambda_{\rm opt}$~  \\[0pt] 
 & &  ($\lambda=0$) & ($\lambda=\lambda_{\rm opt}$) &  \\[5pt] \hline
 0.01 & 0.01 & 0.05 & 0.0467 & --0.2679  \\
 0.01 & --0.30 & 0.0375 & 0.0375 & 0.0111   \\
 0.20 & 0.01 & 5.0375 & 5.0251 & --0.0498 \\ 
0.20 & 0.20 & 0.05 & 0.0467 & --0.2679 \\ 
 0.20 & --0.30 & 0.0431 & 0.0414 & 0.2014 \\
 --0.30 & 0.01 & 11.288 & 11.275 & 0.0333 \\
 --0.30 & 0.20 & 0.0656 & 0.0597 & 0.3139 \\
--0.30 & --0.30 & 0.05 & 0.0467 & -0.2679 \\
\end{tabular}
\end{ruledtabular}
\caption{The sum of the variances $\varf\varphi_y + \varf\varphi_z$ for selected values of $\varphi_y$ and $\varphi_z$ with $(\Delta \hat{n})^2 =10$, both when no entanglement is present between the gyroscopes ($\lambda=0$) and when the optimal probe state is used ($\lambda=\lambda_{\rm opt}$). Here, $\theta = 0.02$ rad, and $\delta = 0.013$ rad. 
\label{tab:stats}}
\end{table}

\subsection{Correlation coefficients for misaligned gyroscopes}\noindent
When alignment errors are introduced, the situation is much more complicated. Fig.~\ref{fig:crb}a shows the dependence of the parameter errors as a function of $(\Delta \hat{n})^2$. Fig.~\ref{fig:crb}b indicates that, for the values of $\varphi_y$ and $\varphi_z$ shown, the combination $\varf\varphi_y+\varf\varphi_z$ is optimal for some non-zero value of the correlation coefficient $\lambda$, whilst the minimum mean squared errors in the nuisance parameters, $\theta$ and $\delta$, are close to zero correlation for this example. This demonstrates that entanglement can be beneficial in some circumstances, at least for a subset of the unknown quantities, and the optimum correlation is not necessarily the same for all of the parameters. We therefore need to study this behavior in more detail. 

When operating as a gyroscope, the input states should be selected to minimize the variance of the measured Sagnac phases, $\varphi_y$ and $\varphi_z$, thereby improving the accuracy of the rotation rates measured by the sensor. This is the asymmetric condition shown in Fig.~\ref{fig:crb}b, minimizing the combination $\varf\varphi_y+\varf\varphi_z$ to find the optimum correlation coefficient $\lambda_{\rm opt}$. Some example values are shown in Table~\ref{tab:stats}. The variances depend strongly on the actual values of $\varphi_y$ and $\varphi_z$ but the gain in precision by choosing $\lambda_{\rm opt}$ instead of $\lambda=0$ appears to be modest for these examples. The optimal states for a given parameter can be found by constructing equal superpositions of the eigenstates with minimum and maximum eigenvalues of the corresponding generator of translations $G_j$ in Eq.~(\ref{eq:generators}) \cite{giovannetti11}. It is generally difficult to create these optimal states, and in addition we require a strategy to choose between different probe states optimized for different parameters. Given the modest improvement in precision it is questionable whether the extra effort would be merited. The typical reduction in the variances shown in Table \ref{tab:stats} is of the order of 5-10\%. However, it is important to remember that the accuracy of gyroscopes in a strapdown inertial navigation affects the accuracy of all of the derived quantities, because the orientation information provided by the gyroscopes is used to resolve the measured accelerations into a set of reference axes to then determine the velocity and position information. This sensitivity to errors in orientation can mean that even marginal gains in the accuracy of the gyroscopes may be important for the overall performance of the system. The trade-off between the difficulty in preparing the entangled states and the benefits of using these states will therefore be application specific.

\begin{figure}[t]
\includegraphics[width=8.5cm]{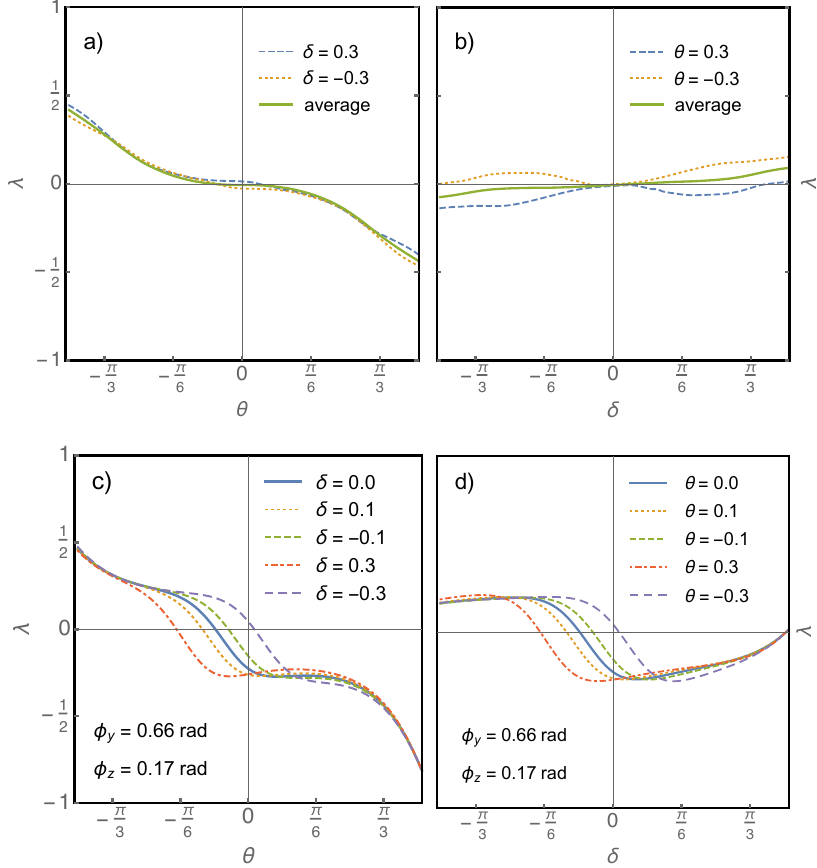}
\caption{Color online. The correlation coefficient $\lambda$ that produces the lowest joint variance $\varf\varphi_y+\varf\varphi_z$ for $(\Delta \hat{n})^2 =10$ as a function of a) $\theta$ and b) $\delta$, both averaged over all values for the phases, for different values of $\delta$ and $\theta$ respectively. The averaged values are calculated by averaging over values of $\delta,\theta=\pm 0.01$, $\pm 0.1$, $\pm 0.2$ and $\pm 0.3$  rad. c) and d) give the same calculations for specific values of the measured phases, $\varphi_y=0.66$ rad and $\varphi_z=0.17$ rad, and the nuisance parameters.}
\label{fig:lambda}
\end{figure}

The case that we are considering here is not the general operation of a gyroscope, it is the situation where the gyroscope is being calibrated to estimate the misalignment of the individual sensors, using the multi-position test described above. In cases where the phases are completely unknown initially, a reasonable approach is to examine the behavior of this optimum value of $\lambda$ as a function of $\theta$ and $\delta$ when averaged over all possible values for $\varphi_y$ and $\varphi_z$. These results are shown in Fig.~\ref{fig:lambda}a and Fig.~\ref{fig:lambda}b, minimizing the errors in the measured phases $\varf\varphi_y + \varf\varphi_z$ as before. The optimum value of $\lambda$ for a wide range of the nuisance parameters is very close to zero ($-\frac{\pi}{6}< \theta,\delta <\frac{\pi}{6}$), when averaged. In particular, the optimum value is zero for small values of $\theta$ and $\delta$, and the response of each is only very weakly dependent on the value of the other nuisance parameter. (A non-zero value for $\lambda$ would indicate that entanglement is beneficial in improving the overall precision of the optical gyroscope). This implies that, in the absence of information regarding the measured phases, the calibration of small nuisance parameters alone will be hindered by the presence of correlated input states. It is only when the coupling between the two gyroscopes becomes significant that correlated input states and entanglement could be beneficial. This is in contrast to the case where the measured phases are known initially, Fig.~\ref{fig:lambda}c and Fig.~\ref{fig:lambda}d, where the optimal value of the correlation coefficient is given as a function of the nuisance parameters for $\varphi_y=0.66$ rad and $\varphi_z=0.17$ rad. In the cases shown, the optimum value for $\lambda$ is strongly dependent on $\theta$ and $\delta$. There is clearly a complicated relationship between the optimal degree of correlation between the two Sagnac interferometers, which varies significantly not only with $\theta$ and $\delta$, but also with $\varphi_y$ and $\varphi_z$.

\begin{figure}[t]
\includegraphics[width=8.5cm]{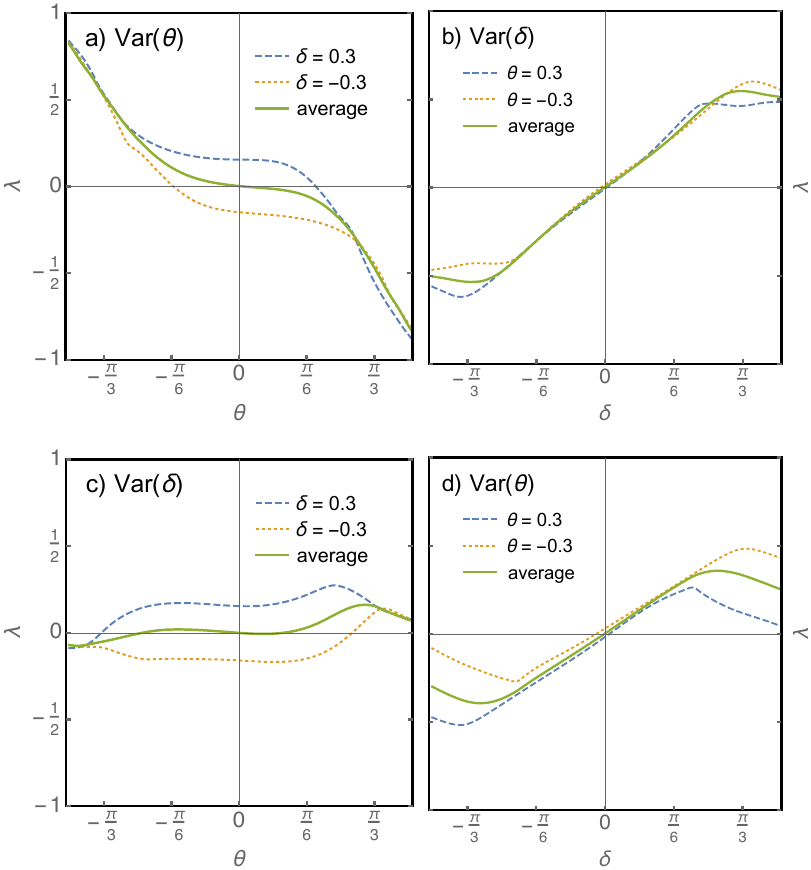} 
\caption{Color online. The correlation coefficient $\lambda$ that minimizes the variance in the estimates of the nuisance parameters for $(\Delta \hat{n})^2 =10$, averaged over all values for the phases: a) $\lambda$ as a function of $\theta$ minimizing $\varf\theta$, b) $\lambda$ as a function of $\delta$ minimizing $\varf\delta$, a) $\lambda$ as a function of $\theta$ minimizing $\varf\delta$, and a) $\lambda$ as a function of $\delta$ minimizing $\varf\theta$---each for several values of the other nuisance parameter and when averaged over values of $\delta,\theta= \pm 0.01$, $\pm 0.1$, $\pm 0.2$ and $\pm 0.3$ rad.}
\label{fig:lambda2}
\end{figure}

Starting with unknown values for the measured phases, the results shown in Fig.~\ref{fig:lambda}a and Fig.~\ref{fig:lambda}b indicate that correlated inputs will not necessarily help to improve the accuracy of the estimates of the nuisance parameters, for small values of $\theta$ and $\delta$ at least. However, as the estimates of the Sagnac phases $\varphi_y$ and $\varphi_z$ improve, the results shown in Fig.~\ref{fig:lambda}c and Fig.~\ref{fig:lambda}d indicate that the optimum value for $\lambda$ becomes strongly dependent on the values of the nuisance parameters. This implies that the optimal calibration process could involve an adaptive approach, with the selected value of $\lambda$ being dependent on the estimated values and expected errors in $\varphi_y$ and $\varphi_z$ and the estimated nuisance parameters themselves. The problem becomes even more complicated when one takes into consideration that the optimum value of $\lambda$ calculated in Fig.~\ref{fig:lambda} is the value that minimizes the combination $\varf\varphi_y+\varf\varphi_z$, i.e. the variance in the phase measurements, not the variance in the nuisance parameters. Fig.~\ref{fig:lambda2} shows the behavior of the optimum correlation coefficient when minimizing the variance in the estimated value for each of the nuisance parameters, again averaged over the measured phases. 

\begin{figure}[t]
\includegraphics[width=8.5cm]{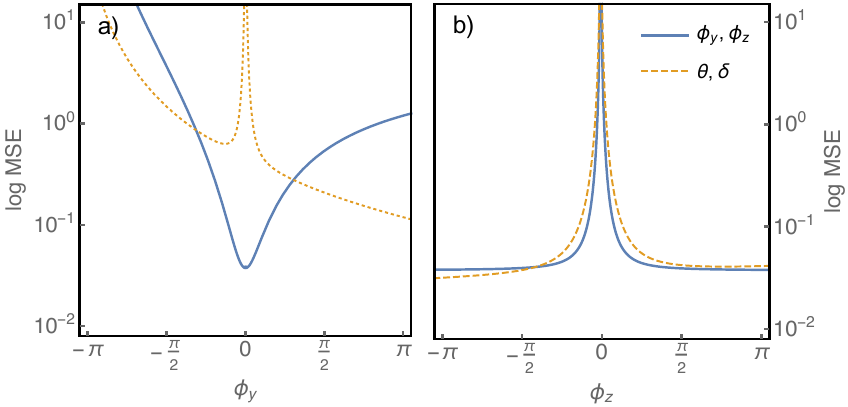} 
\caption{Color online. The mean square errors for $\bm{\varphi}$ as functions of a) $\varphi_y$ with $\varphi_z = 0.17$~rad, and b) $\varphi_z$ with $\varphi_y = 0.66$~rad, for $(\Delta \hat{n})^2 =10$, $\theta = 0.02$~rad, and $\delta = 0.013$~rad. The blue solid line indicates the combined log MSE in $\varphi_y$ and $\varphi_z$, while the orange dotted line indicates the combined log MSE in $\theta$ and $\delta$. Calibration using specially selected $\varphi_y$ and $\varphi_z$ achieve the best results when adaptive strategies are used, i.e. when the values of $\varphi_y$ and $\varphi_z$ are chosen based on the inferred values of $\theta$ and $\delta$.}
\label{fig:phi}
\end{figure}

Whilst the different optimisation criteria shown in Fig.~\ref{fig:lambda2} do show qualitative and quantitative differences when compared with the results shown in Fig.~\ref{fig:lambda}a and Fig.~\ref{fig:lambda}b, the results are not {\em inconsistent}---meaning that the optimal values for $\lambda$ are zero or close to zero in each case, i.e. in the physically relevant regime where the misalignment parameters are small. It is only when $\delta$ is relatively large ($|\delta| \simeq 0.3$ rad) that the optimal value for the correlation is non-zero at $\theta\simeq 0.0$. This means that an adaptive calibration method should be feasible, although the precise form that it should take will depend on the criteria being optimized and the quality of the a priori knowledge regarding the rotation rates applied to the system, and the corresponding Sagnac phase measurements. Ideally, the calibration would be performed using a rotation rate that had been selected to produce a minimum variance for the parameter estimates, and these vary significantly, as can be seen in Table~\ref{tab:stats}. Fig.~\ref{fig:phi} shows the mean square errors in the four estimated parameters as functions of the applied Sagnac phases. 

\subsection{3D gyroscopes}\noindent
In our discussion so far we have ignored the third dimension. Here we give a simple argument that shows that the calibration process for quantum interferometers can be generalized to three dimensions in a straightforward way. Consider the three axes determined by three Sagnac gyroscopes. We pick one axis as a reference axis---such as the $y$-axis in the discussion above. The remaining two axes deviate from a perfect triad in three angles, namely the angles for the $x$ and $z$-axes with the $y$-axis, and the angle between the $x$-axis and the $z$-axis. These are three nuisance parameters $\bm\theta$ (now a three-dimensional vector) arising from the physical tolerances inherent in any manufacturing process. Calibration requires that we determine these angles with high precision. During operation, any Sagnac phase measurements can be corrected by removing the coupling introduced by the nuisance parameters in post-processing. 

The calibration consists of applying a well-defined rotation rate to the system, fixed with respect to an external frame, which leads to expected values of the phases $\varphi_x$, $\varphi_y$, and $\varphi_z$. We must estimate these phases with the gyroscope. We then rotate the gyroscope sensor through a known angle and repeat the measurements with the same rotation rate applied. For each independent set of phase measurements we take three readings, namely the photon number difference in $x$, $y$, and $z$. The question is now if we can take enough \emph{different} readings by applying these known rotations to the three-dimensional gyroscope several times, taking into account that each finite rotation is itself associated with a nuisance parameter, such as the parameter $\delta$ above. If we take $K$ sets of readings, after each rotation about an arbitrary angle, $\bm{\delta}_j$ ($1\leq j\leq K$), we introduce---in principle---three new nuisance parameters for each $\bm\delta_j$. It is easy to see that the total number of parameters stays ahead of the total number of readings by three, and we will not be able to establish all parameter values unambiguously. To overcome this problem, we rotate the gyroscope around the same externally defined axis that determine the calibration signal $(\varphi_x, \varphi_y, \varphi_z)$. This will incur only a single nuisance parameter for every rotation (after the first set of measurements), namely the rotation angle, and it still allows us to rotate the gyroscope in any direction we wish. Suppose we take a set of measurements in $K$ gyroscope orientations. We will have the three components of the applied rotation rate, three angles corresponding to the mis-alignment of the Sagnac axes, plus $K-1$ rotation angles, giving a total of $K + 5$ parameters that need to be estimated from $3K$ measurements of the Sagnac phases. A successful calibration therefore requires 
\begin{align}
3K \geq K+5\, ,
\end{align}
which means that we need at least three sets of measurements. If we then wish to relate the gyroscope orientation to an external reference frame we need another three parameters, and this can be achieved by taking another set of measurements at a fourth orientation. This analysis implies that the generalisation of our analysis to three dimensions does not pose a major obstacle---which is to be expected since similar processes are already used in inertial navigation systems and can include more complicated sequences of rotations and applied rotation rates to characterize other errors~\cite{savage98,aggarwal08}---and there is no reason to believe that the main results given above for entangled input states would simplify significantly in moving from two to three dimensions. If anything, the optimisation of the input states in three dimensions would be expected to be an even more complicated problem than the case presented here.

\section{Conclusions} \noindent
We have studied the role of entanglement in the calibration and operation of an optical quantum gyroscope, relying on measurements of the Sagnac phase shift. The accuracy of the gyroscope is an important factor in inertial navigation systems, and any imperfections in the system must be well characterized for reliable operation as a rotation sensor. The interferometers are sensitive to rotation rates, which must be integrated with respect to time to generate orientation information, which can then be used to resolve acceleration measurements into a reference frame, thereby allowing changes in velocity and position to be determined. 

For a perfectly aligned gyroscope, the optimal state in each individual interferometer will be highly entangled, NOON states for example, with no entanglement or other correlations between the interferometers. The Sagnac phase measurements are essentially independent, and the quantum Cram\'er-Rao bound can be achieved asymptotically. When the  interferometers are not perfectly orthogonal to each other, the appearance of nuisance parameters (the mis-alignment of the sensors) will give rise to couplings between the measured values for the rotation rates. Knowing the values of these nuisance parameters allows the effect of these couplings to be reduced or removed, thereby improving the accuracy of the system. To this end, the system must be calibrated by taking readings from the gyroscope in a variety of known orientations. In two dimensions, we have demonstrated that the optimal states for such systems are dependent on the values of the nuisance parameters and the condition that is being optimized. Specifically, the states that would be optimal for the operation of the gyroscope as a sensor---i.e. the states that minimize the variance of the phase measurements---can be different to states that optimize the calibration of the nuisance parameters. In fact, the calibration of the optical quantum gyroscope is complicated by the fact that the variances of our estimators depend strongly on the values of the various parameters in the problem. In addition, we have found that the precision of the measurements of both the nuisance parameters and the applied signal is highly dependent on the amount of entanglement, as indicated by the correlation coefficient $\lambda$. These dependencies must be fully understood, as the use of suboptimal probe states can make the calibration worse. Small improvements in the estimation of the nuisance parameters may seem to come at the expense of hard to engineer quantum probe states, but the nature of error accumulation in inertial navigation means that even modest gains in accuracy can lead to an important operational improvement of slower error divergence. 

An important open question is how to establish the optimal quantum strategy for the calibration of the optical quantum gyroscope: what are the optimal quantum states to send into the three interferometers, and how should we divide our resources between the different probe states? Even when the measured observables commute, it is not immediately clear how the probe states must be chosen in order to maximise the information gain about the different parameters in the gyroscope. These questions will be addressed in future work.

\section*{Acknowledgments}\noindent
PK and JFR would like to thank the Isaac Newton Institute for Mathematical Sciences, Cambridge, for support and hospitality during the Quantum Control Engineering programme (July-August 2014), where part of this work was undertaken. PK acknowledges EPSRC for funding via the Quantum Communications Hub, JD acknowledges EPSRC for funding via the NQIT Hub, JFR acknowledges EPSRC for funding via the Quantum Technology Hub in Sensors and Metrology.

\begin{widetext}
\appendix

\section{Explicit forms for $F_k$}\label{app:F}\noindent
Here we give the explicit forms for $F_k$ in Eq.~\eqref{eq:qfifinal} in the case where $ (\Delta \hat{n}_y)^2 = (\Delta \hat{n}_z)^2 \equiv (\Delta \hat{n})^2$ and $C_{yz} = \lambda (\Delta \hat{n})^2$. For notational compactness, let $\gamma = \beta(\theta+\delta+\pi/2)$, $\beta_\theta \equiv \beta(\theta)$, and $\beta_\delta \equiv \beta(\delta)$. The functions $F_k$ then become
\begin{align}
 F_y & =  \left[ \gamma(\beta_\delta+\beta_\theta) \cos\theta-\beta_\delta\beta_\theta\sin (\delta +\theta ) \right]^2 = (\det M)^2\, , \\
 F_z & =  \gamma^2 \left[ \beta_\delta^2\left(2 \lambda  \sin\theta+\sin ^2\theta +1\right) +2 \beta_\delta \beta_\theta \sin\delta (\sin\theta+\lambda ) + \beta_\theta^2 \left(\sin ^2\delta+1\right)\right] +  \beta_\delta^2 \beta_\theta^2\left[\cos ^2(\delta +\theta )+1\right] \\
 & \quad- 2 \gamma \beta_\delta \beta_\theta  \left\{ \beta_\delta \cos (\delta +\theta ) (\sin\theta+\lambda )+\beta_\theta \left[\sin\delta \cos (\delta +\theta )+\lambda \right]\right\} , \cr
 F_\theta & =  \gamma^2\cos ^2\theta \left[(\sin\delta-\sin\theta)(\sin\delta-\sin\theta-2 \lambda )+2\right]  \\
 & \quad + \beta_\delta^2 \left[ \frac32 - \lambda  \sin (2 \delta +3 \theta )+\sin\delta \sin (\delta +2 \theta )+\frac12\cos (2\delta +4 \theta ) + \lambda  \sin\theta \right] \cr 
 & \quad - \gamma \beta_\delta \cos\theta \left\{ \lambda  \left[ \cos (2 \delta +\theta )-3 \cos (\delta +2 \theta )+\cos\delta+\cos\theta \right]+3 \sin (\delta +\theta )+\sin (2\delta +2\theta)-\sin (\delta +3 \theta )-\sin 2 \theta \right\} \, , \cr
 F_\delta & =  \gamma^2\cos ^2\theta \left[ (\sin\delta-\sin\theta) \left(\sin\theta -\sin\delta+2 \lambda \right)-2\right] \\
 & \quad-\beta_\theta^2 \left\{\left(\sin ^2\delta+1\right) \sin ^2(\delta +\theta )+\cos ^2\theta \left[\cos ^2(\delta +\theta )+1\right]-2 \cos\theta \sin (\delta +\theta ) \left[\sin\delta \cos (\delta +\theta )+\lambda \right]\right\} \cr
 & \quad +\gamma\cos\theta \beta_\theta [\lambda  (\cos (2 \delta +\theta )+\cos (\delta +2 \theta )+\cos\delta-3 \cos\theta)+3 \sin (\delta +\theta )+\cos(2\theta+2\delta)\sin(\theta-\delta) - \sin 2 \delta ] \, . \nonumber
\end{align}
\end{widetext}

\end{document}